\documentclass[superscriptaddress,showpacs,twocolumn]{revtex4}
\usepackage{graphicx}

\begin{document}

\title{A simple and robust single-pixel computational ghost imaging}

\author{Lijun Song}
\affiliation{Department of Applied Physics, Changchun University,
Changchun 130022, China}
\author{Cheng Zhou}
\affiliation{School of Science, Changchun University of  Science and Technology,
Changchun 130022, China}
\author{Li Chen}
\affiliation{Department of Applied Physics, Changchun University,
Changchun 130022, China}
\author{Xiaoguang Wang\footnote{E-mail: xgwang@zimp.zju.edu.cn }}
\affiliation{Zhejiang Institute of Modern Physics, Department of Physics, Zhejiang University, Hangzhou 310027, China}
\author{Jing Cheng\footnote{E-mail: phjcheng@scut.edu.cn }}
\affiliation{Department of Physics, South China University of Technology, Guangzhou 510640, China}
\begin{abstract}

A simple and robust experiment demonstrating computational ghost imaging with structured illumination and a single-pixel detector has been performed. Our
experimental setup utilizes a general computer for generating pseudo-randomly patterns on the liquid crystal display screen to illuminate a
partially-transmissive object. With an incoherent light source, this object is imaged. The effects of light source, light path, and the number of measurements on the reconstruction quality of the object are discussed both theoretically and experimentally. The realization of computational ghost imaging with computer liquid crystal display is a further setup toward the practical application of ghost imaging with ordinary incoherent light.

\end{abstract}

\maketitle

\section{Introduction}

Ghost imaging (GI) is a promising imaging technique based on the classical or quantum correlation of the light field fluctuations, which can realize
the reconstruction of an object by means of intensity correlation of two light beams, i.e., the object beam and the reference beam. The first GI
experiment was performed by using two-photon entangled light generated in spontaneous parametric down-conversion \cite{1995}. Subsequently a lot of
works are focused on GI with thermal sources because it may be useful in practical application \cite{bennink,gatti,chengji1,chao,ferri,valencia,cai,wlan,BJJ}. In recent years, a
significant number of theoretical and experimental schemes about GI have been presented, such as computational GI \cite{g4}, compressive GI \cite{katz}, differential GI \cite{ferri1,li}, and high-order GI \cite{chen,kmr}. Among them, computational GI was first proposed theoretically by Shapiro \cite{g4}. Compared with the standard pseudothermal GI, one advantage of computational GI is that only a simple detector of single-pixel resolution is needed in the test path and replaces high spatial-resolution detector of the reference beam with a computation of the propagating field. Computational GI leads to single-pixel imaging employing structured illumination, which simplifies the implementation of an imaging system.

Silberberg \emph{et al.} first achieved computational GI in experiment by using only a single detector, where the essence is to replace the rotating diffuser with a computer-controlled spatial light modulator (SLM) \cite{g5}. Then, in some computational GI schemes, a digital micromirror device (DMD) is used to generate random spatial distribution as an SLM  \cite{D1,D2,cs,sun1,sun2}. Recently, Sun \emph{et al.} demonstrated three-dimensional \cite{sun1} and full-color \cite{sun2} computational imaging by replacing the SLM and laser with a digital light projector (DLP) and using several single-pixel detectors in different locations. This technique has
attracted great interest \cite{g7}. But until recently, an SLM or DMD or DLP is necessary in the computational GI schemes. These devices are costly or difficult to modulate. Meanwhile, the major drawbacks of computational GI include having to integrate and average over thousands of frames, requiring large computer storage space, long processing time and fast detectors. Due to GI's significant application value, it is getting more and more attention. Hence, how to improve the image quality with simple and practical devices becomes the research focus.

In this paper, we proposed a new experimental scheme for computational GI, where structured illumination and a single-pixel detector are used. A general liquid crystal display (LCD) is used as light source, the pseudo-randomly patterns directly generated by the computer on the display do structured illumination, and this computer performs the reconstructions of the test object. This greatly simplifies the control process of light source in computational GI, and there is no need to modulate SLM by programming. This structured illumination can greatly reduce the number of measurements required for a faithful reconstruction.
In the meantime, the experimental setup in this work is very simple and has excellent robustness to external light sources of noise.
Therefore, our current work offers great potential for future implementations of GI in practical applications.

\section{experimental setup}

\begin{figure}[htbp]
\includegraphics[width=3in]{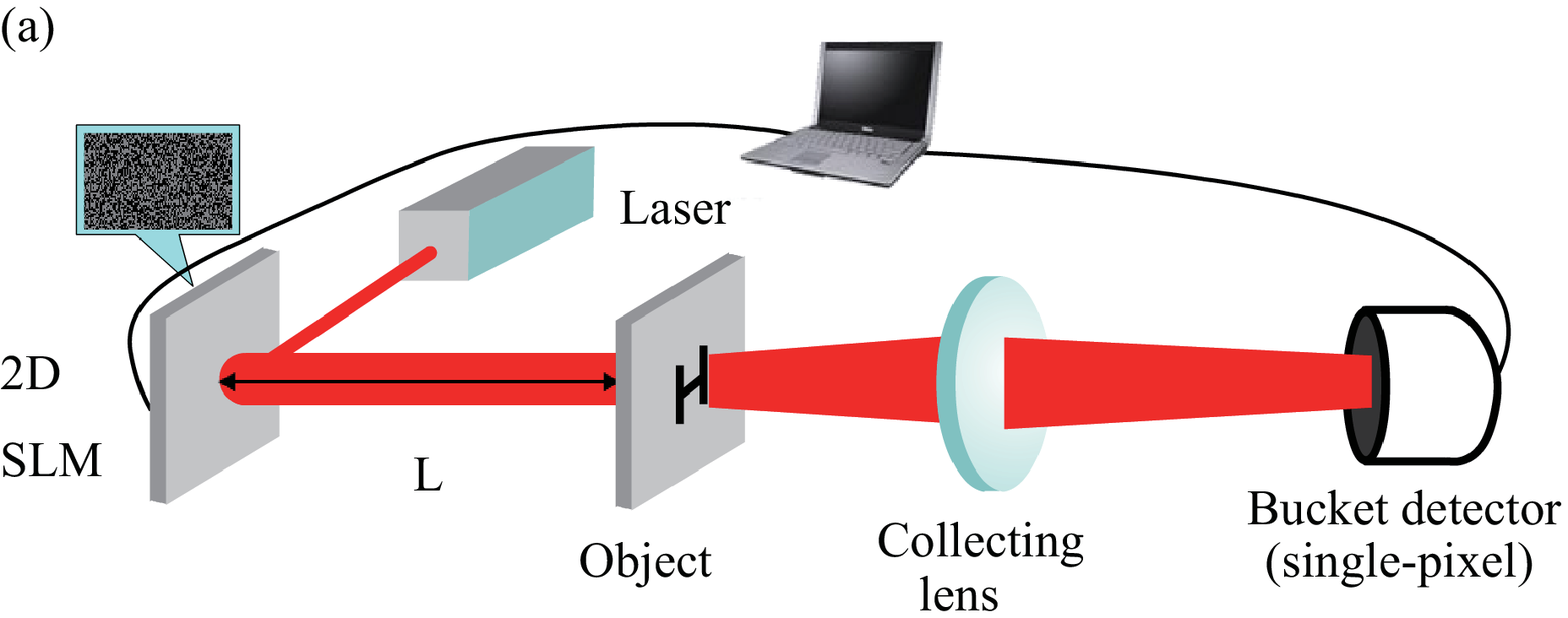}
\includegraphics[width=3in]{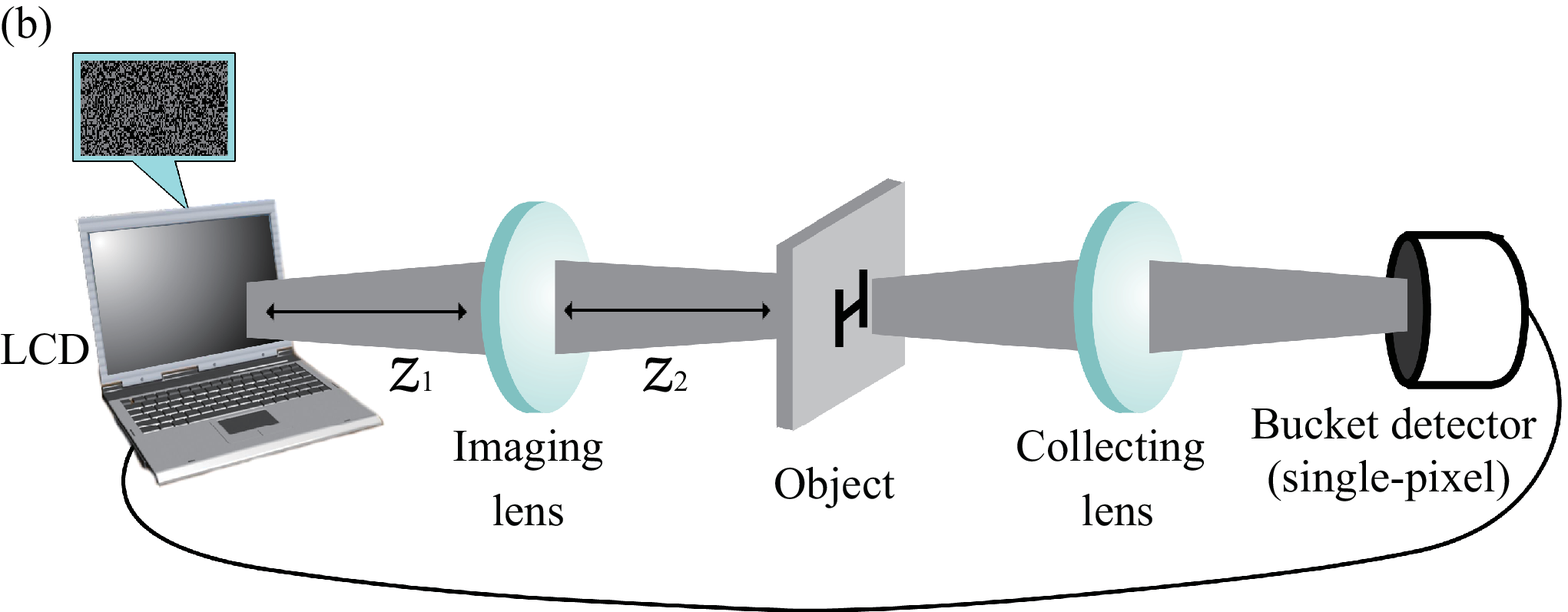}
\caption{(Color online) Experimental setups for computational GI. (a) Standard computational GI experimental setup.
(b) The computational GI setup with computer LCD used in this work. The LCD illuminates the
object with computer-generated random speckle patterns. The light transmitted through the
object is collected onto the single-pixel photodetector.}
\label{1}
\end{figure}

This experimental setup is shown in Figure~\ref{1}. In traditional computational GI, a computer-controlled SLM is utilized, which
works as a controlled phase mask for the spatial phase of the light field [Fig.~\ref{1}(a)]. A pseudothermal light beam  is generated by applying phase patterns on a SLM irradiated by laser \cite{g5}. A spatially incoherent light beam is generated by applying pseudo-random phase patterns on a SLM.
Our ghost imaging system replaces the laser and SLM with a general computer LCD [Fig.~\ref{1}(b)]. The optical setup can be divided into two parts: the illumination system and the collection system, located before and behind of the object, respectively. In the illumination system, the LCD is used as light source, the pseudo-randomly patterns are directly generated by the computer on the LCD screen. Suppose the light source (here the LCD screen) is located at $z=0$, an imaging lens ($f=15~\rm cm$, $R=15~\rm mm$) is located at $z=z_1$ , an object `H' with the size of $1.0\times0.7~{\rm cm}^2$ is located at $z=z_1+z_2$. The collection system is consists of a lens and a single-pixel detector. This lens is used to collect all transmitted light through the object onto the bucket detector.

Here we use the computer LCD screen (Dell E176FP, 17~\rm inch, Dot pitch 0.264~\rm mm, Screen ratio $5:4$). Without loss of generality, our experimental scheme with other LCD screens can also be achieved. The light source in our experimental setup has a good stability and high imaging quality.
Normally, ghost imaging experiment needs to be done in the dark since all external sources of light would interfere with the recorded voltage from the single-pixel detector during the data acquisition process. Nevertheless, we proved that we can also obtain a clear image of the object even in the presence of external light sources,
which indicates that our system is very robust. To sum up, high-quality reconstruction can be achieved by making use of the simple and robust experimental setup in our scheme.

\section{Theoretical analysis}
We firstly generate a series of patterns in the screen used as the sources. These sources are labeled as $I_{\text{soc}}^{(n)}(x,y),
n=1,2,\cdots,N$. In current experiments, these $I_{\text{soc}}^{(n)}(x,y)$ are generated by a random function in Matlab, so that statistically ($N$ is large enough) $I_{\text{soc}}^{(n)}(x,y)$ should satisfy \cite{g8}
\begin{equation}\label{e1}
\frac{1}{N}\sum\limits_{n=1}^{N}{I_{\text{soc}}^{(n)}(x,y)I_{\text{soc}}^{(n)}({{x}_{0}},{{y}_{0}})}\to C_0+C_1\delta (x-{{x}_{0}},y-{{y}_{0}}),
\end{equation}
where $C_0$ and $C_1$ are two constants. Since the light from different emitters in the computer screen are incoherently distributed, we may consider the imaging lens plays the role of an incoherent imaging system, thus the intensity distribution just before the object (i.e., at $z=z_1+z_2$ ) has the form
\begin{equation}\label{e2}
I_{\text{obj}}^{(n)}(x',y')=\int{dxdyI_{\text{soc}}^{(n)}(x,y){{\left| h(x'-Mx,y'-My)\right|}^{2}}},
\end{equation}
where $M=-z_2/z_1$ is the magnification of the system, and $h(x'-Mx,y'-My)$ is the point spread function (PSF)
\begin{eqnarray}\label{e3}
&&{\left|h(x'-Mx,y'-My)\right|}^{2}\cr\cr
&=&\bigg|\int{dudvP(u,v){e}^{\frac{j\pi}{\lambda}(\frac{1}{z_1}+\frac{1}{z_2}-\frac{1}{f})(u^2+v^2)}}\cr
&&\times e^{-\frac{j2\pi}{\lambda{{z}_{2}}}\big((x'-Mx)u+(y'-My)v\big)}\bigg|^2,
\end{eqnarray}
where $P(u,\nu)$ is the aperture function of the lens, here $P(u,\nu)=1$ for
$u^2+\nu^2\leq R^2$ and $0$ for other values.

Now, the bucket detector will detect a total energy signal
\begin{eqnarray}\label{e40}
{B}^{(n)}&=&\int{dx'dy'I_{\text{obj}}^{(n)}(x',y')T(x',y')},
\end{eqnarray}
where $T(x',y')$ is intensity transmittance.
Substituting Eqs.~(\ref{e2}) and (\ref{e3}) into Eq.~(\ref{e40}), we can obtain
\begin{eqnarray}\label{e4}
{B}^{(n)}&=&\int{dx'dy'T(x',y')}\int{d{{x}_{0}}d{{y}_{0}}I_{\text{soc}}^{(n)}({{x}_{0}},{{y}_{0}})}\cr\cr
&&\times{{\left| h(x'-M{x_0},y'-M{y_0}) \right|}^{2}}.
\end{eqnarray}

The ghost image can be obtained by calculating the intensity correlation between the light sources and the signals of the bucket detector
\begin{eqnarray}\label{e50}
S(x,y)=\frac{1}{N}\sum\limits_{n=1}^{N}{I_{\text{soc}}^{(n)}(x,y){{B}^{(n)}}}.
\end{eqnarray}
Then, substituting Eqs.~(\ref{e1}) and (\ref{e4}) into Eq.~(\ref{e50}), the ghost image is given by
\begin{eqnarray}\label{e5}
S(x,y)&=&\int{d{x}'d{y}'d{{x}_{0}}d{{y}_{0}}\bigg[\frac{1}{N}\sum\limits_{n=1}^{N}{I_{\text{soc}}^{(n)}(x,y)I_{\text{}}^{(n)}({{x}_{0}},{{y}_{0}})}\bigg]}\cr\cr
&&\times T({x}',{y}'){{\left| h({x}'-M{{x}_{0}},{y}'-M{{y}_{0}}) \right|}^{2}} \cr\cr
&=&background\text{+}C_1\int{d{x}'d{y}'T({x}',{y}')}\cr\cr
&&\times{{\left| h({x}'-M{{x}},{y}'-M{{y}}) \right|}^{2}}.
\end{eqnarray}
It is clear to see that the quality of ghost image is determined by the
PSF [Eq.~(\ref{e3})]. When there exists no defocusing, i.e., $\frac{1}{z_1}+\frac{1}{z_2}-\frac{1}{f}=0$ , the PSF is given by the Fourier transform of the
aperture function
\begin{equation}\label{e6}
{\left| h({x}'-M{{x}},{y}'-M{{y}}) \right|}^{2}={\left| \frac{{{J}_{1}}(2\pi R\rho /\lambda {{z}_{2}})}{R\rho /\lambda {{z}_{2}}} \right|}^{2},
\end{equation}
where $\rho=\sqrt{(x^{^{\prime }}-M x)^2+(y^{^{\prime }}-M y)^2}$, and $J_1$
is the 1st order Bessel function. When there exists defocusing, $\frac{1}{z_1}+\frac{1}{z_2}-\frac{1}{f}\neq0$, the PSF will be degraded, and the ghost
imaging quality will be decreased.

\section{EXPERIMENTAL RESULTS}

\begin{figure}[htbp]
\includegraphics[width=2.2in]{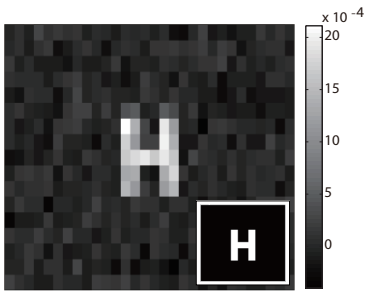}
\caption{Experimental reconstruction of the letter `H' with 4000 measurements using the binary patterns with the speckle size $20\times30$ pixels.
$4000$ randomly distributed binary patterns of $1920\times1080$ pixels with a black-to-white ratio of $1:1$ are projected onto the object. The imaging lens is placed at $z_1=74~{\rm cm}$, and the object is set at a distance $z_2=19~{\rm cm}$
from the imaging lens. The inset indicates the transmission mask. }
\label{experiment}
\end{figure}

The computational GI with a LCD is demonstrated experimentally by constructing the setup shown in Fig.~\ref{1}(b).
A series of binary patterns with the speckle size $20\times30$ pixels are generated by the computer on the LCD screen.
The imaging lens is placed at a distance $z_1=74~{\rm cm}$ from the LCD,
and the transmissive object `H' is placed at a distance $z_2=19~{\rm cm}$ from the imaging lens.
The other lens behind the object collects the transmitted light onto the bucket detector.
The letter `H' is  accurately reconstructed by using 4000 effective measurements, as shown in Fig.~\ref{experiment}.

\begin{figure}[htbp]
\includegraphics[width=3.0in]{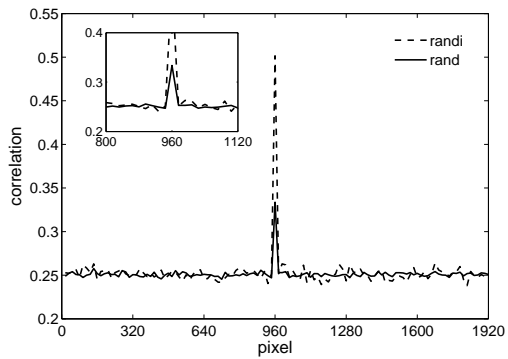}
\caption{Statistical distribution of the light sources. Dashed line corresponds to randomly distributed binary speckle patterns with an equal black-to-white ratio. Solid line corresponds to speckle patterns with the intensity randomly distributed in
$(0,1)$.}
\label{2}
\end{figure}

In the following, we will discuss the possible factors affecting the experiment results in detail.
First, we try to consider the effect of light sources on the quality of the ghost images.
Due to the algorithms and the fluctuations in Matlab functions, statistical distribution of the light sources may be different,
which will further affect the image quality.
We generate $N=4000$ random speckle patterns of $1920\times1080$ pixels
by using the Matlab functions `rand' and `randi', respectively.
The correlation property of light sources is simulated in one-dimensional case, as shown in Fig.~\ref{2}.
The simulation results are consistent with the theoretical prediction [Eq.~(\ref{e1})].
Computation GI experiments with both kinds of pseudo-random speckle patterns are achieved, the results indicate that
there is a slight difference for the quality of the reconstructions.
In our scheme, a series of pseudo-random binary patterns are utilized.

\begin{figure}[htbp]\vspace{1em}
\includegraphics[width=1.5in]{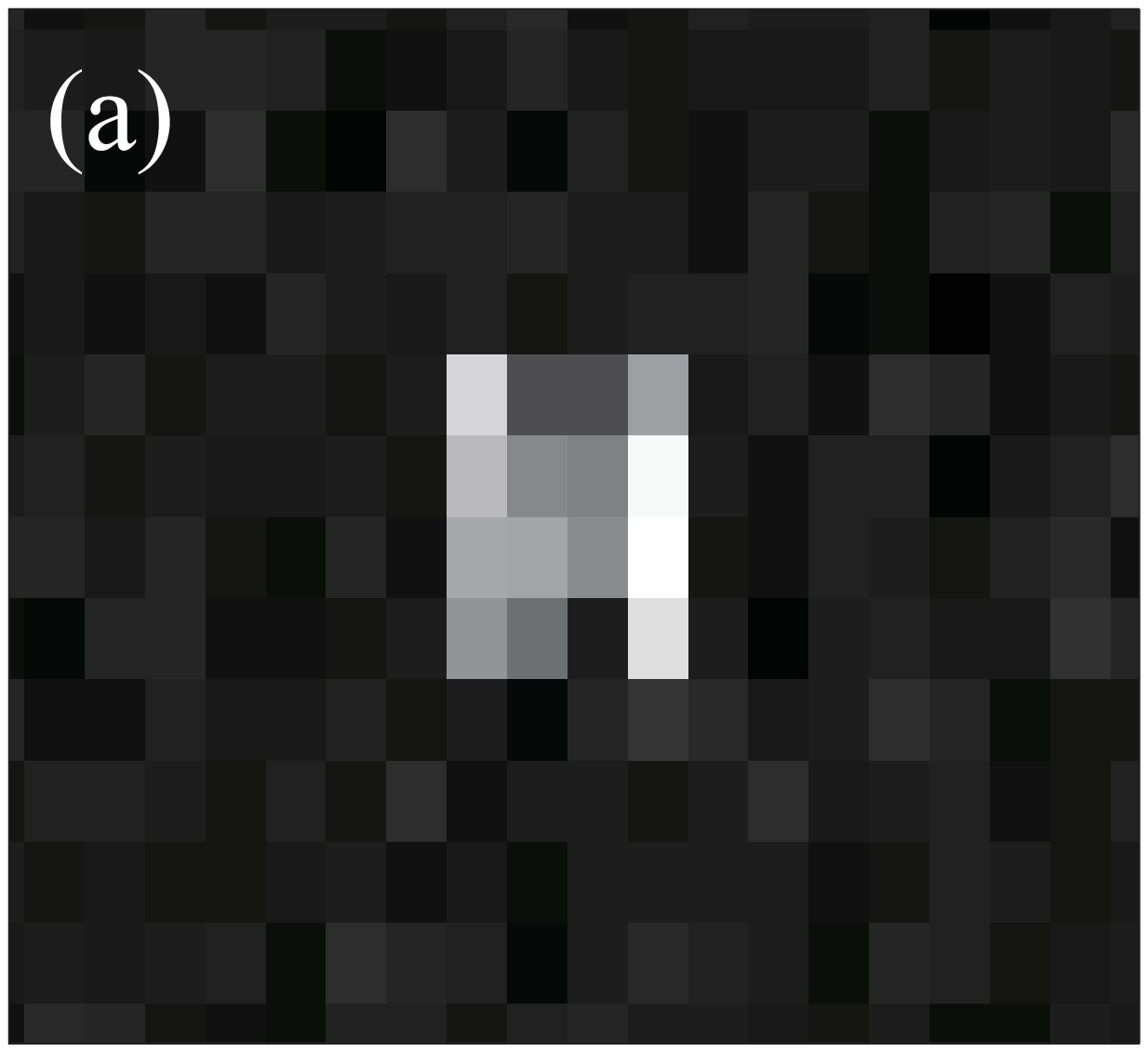}
\includegraphics[width=1.5in]{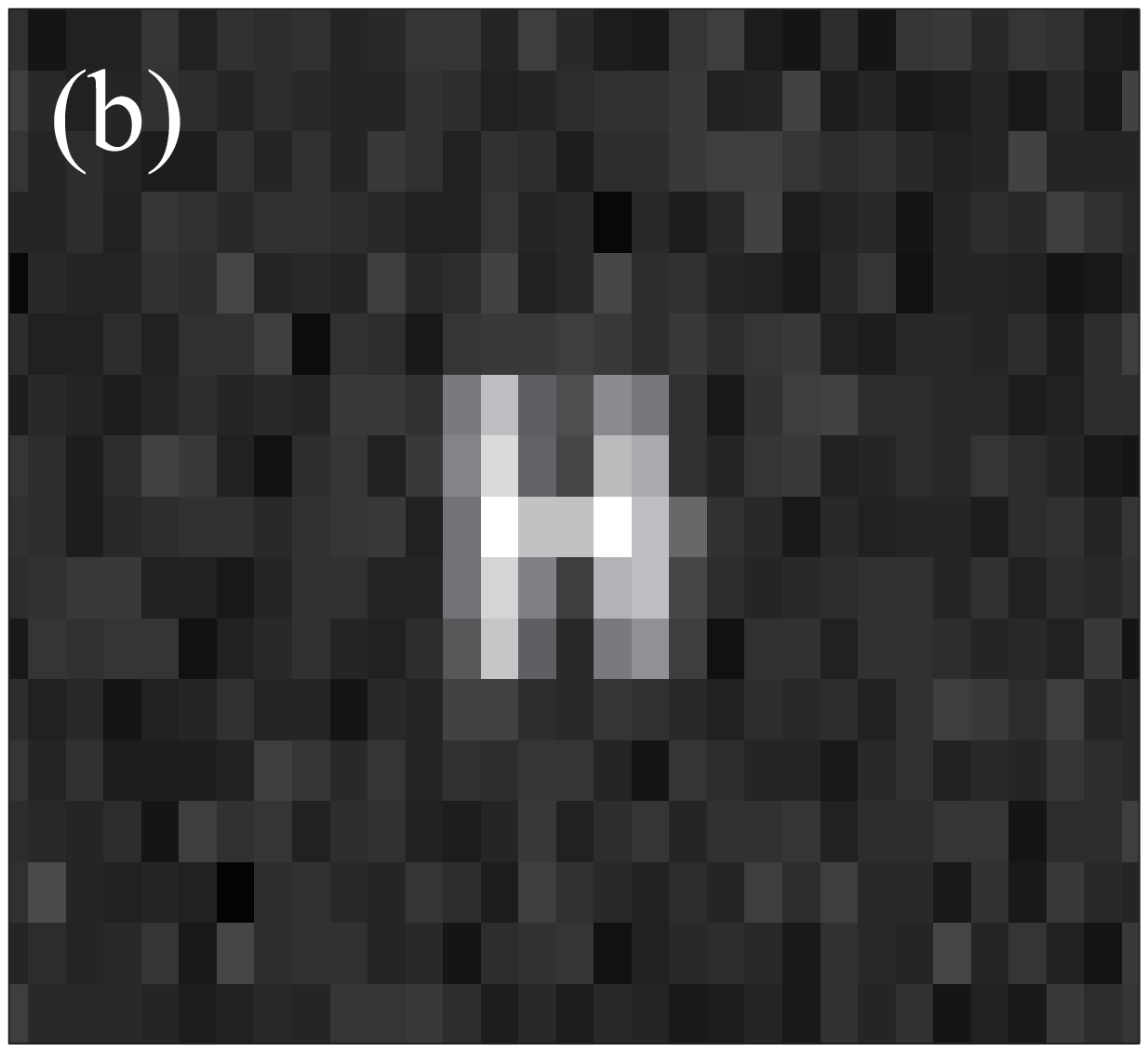}
\caption{Reconstructed images experimentally with 4000 measurements. (a) The patterns with the speckle size $32\times40$ pixels are used. (b) An out-of-focus image reconstructed at a different position $z_2=23~{\rm cm}$. Other experimental parameters are the same as Fig.~\ref{experiment}.}
\label{3}
\end{figure}

In addition, the speckle size may be an important factor to affect image quality.
We firstly generate 4000 speckle patterns with the speckle size $32\times40$ pixels using the computer.
When the number of measurements is 4000, the reconstructed image is displayed in Fig.~\ref{3}(a). The imaging result is not desirable.
If we adjust the speckle size, and use the patterns with the smaller speckles, such as $20\times30$ pixels.
Under the same experimental conditions, the object is reconstructed, as shown in Fig.~\ref{experiment}.
It is evident that the reconstruction by the patterns with the smaller speckles is much clearer by comparing Figs.~\ref{experiment} and \ref{3}(a).
The experimental results show that changing the size of speckles is an effective method to improve imaging quality.
The high-quality image can be obtained by making use of the patterns with the appropriate speckle size.

Second, the influence of light path on reconstruction quality is also discussed.
In Fig.~\ref{experiment}, the object `H' is at a distance $z_2=19~{\rm cm}$ from the imaging lens, where the light path satisfies focusing,
i.e., $\frac{1}{z_1}+\frac{1}{z_2}=\frac{1}{f}$, due to $z_1=74~{\rm cm}$ and $f=15~{\rm cm}$. The accurate reconstruction of the object transmission $T(x',y')$ is achieved by using 4000 measurements [Fig.~\ref{experiment}].
And then we move the object `H' to  a new position at a distance $z_2=23~{\rm cm}$ from the imaging lens so that
 the light path arises the defocusing case,  $\frac{1}{z_1}+\frac{1}{z_2}\neq\frac{1}{f}$,
reconstruction of an image results in an out-of-focus image of the object, as shown in Fig.~\ref{3}(b), which indicates the depth-resolving
imaging capabilities of the computational GI technique in this work.

\begin{figure}[htbp]\vspace{1em}
\includegraphics[width=1.2in]{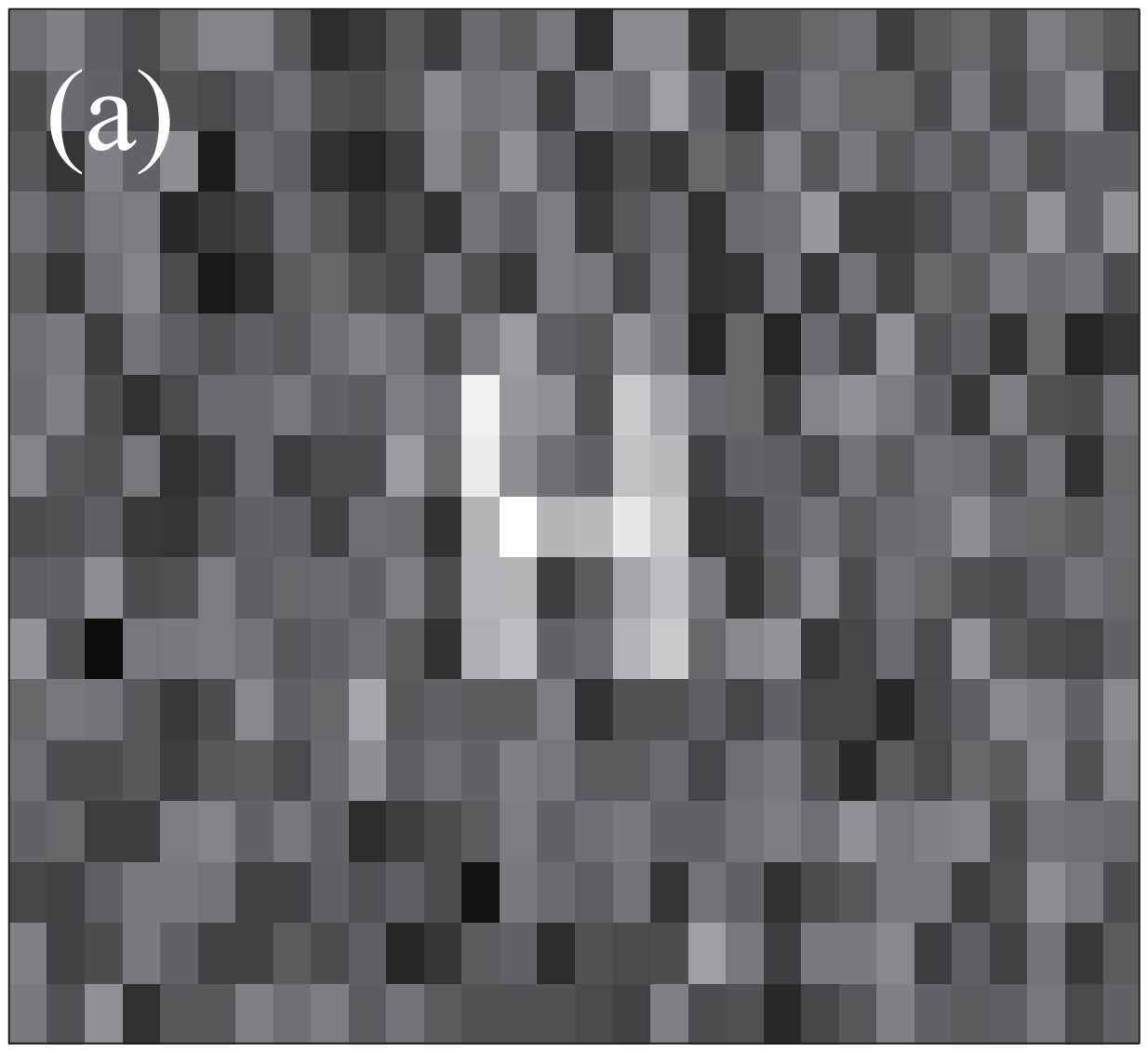}\hspace{-1.2em}
\includegraphics[width=1.2in]{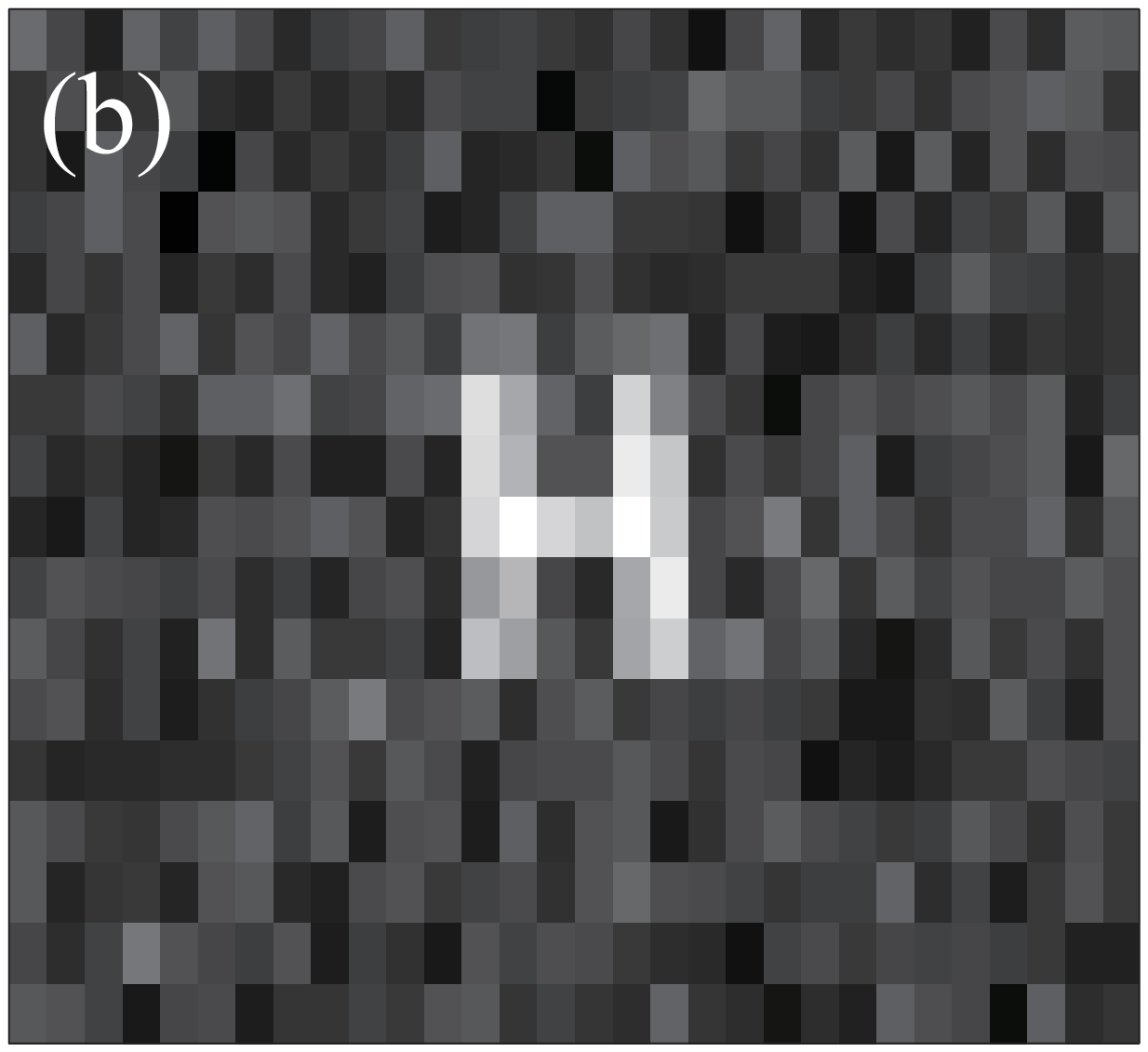}\hspace{-1.2em}
\includegraphics[width=1.2in]{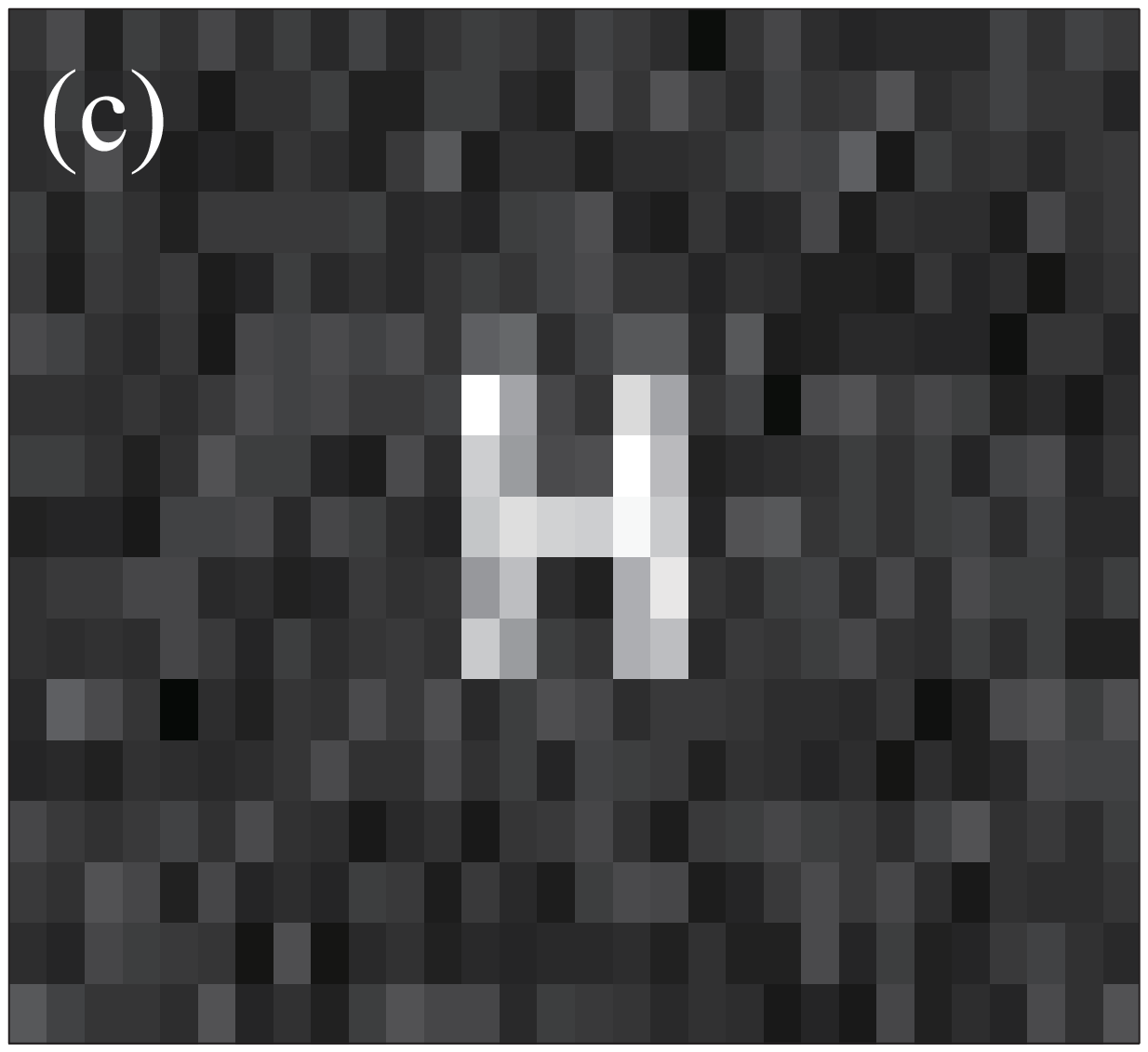}
\caption{Reconstructed
images experimentally for different number of measurements $N$. (a) $N=1000$ (b) $N=2000$ (c) $N=3000$. Other experimental parameters are the same as Fig.~\ref{experiment}.}
\label{5}
\end{figure}

Finally, we investigate the relation between imaging quality and the number of measurements, displaying reconstructed images for different measurements $N=1000,\ 2000$, and $3000$ in Fig.~\ref{5}.
It is easy to observe that as the number of measurements $N$ increases, the quality of reconstructed images enhances significantly.
Furthermore, the use of patterns with the appropriate speckle size can reduce the number of measurements needed for image reconstruction,
so we can obtain high-quality images using only a small amount of data.

\section{summary}

In conclusion, we have demonstrated that computational GI can be
accomplished with pseudo-randomly distributed patterns in the computer
LCD. The high-quality object image can be reconstructed only by a set of simple and low-cost experimental
apparatuses consisting of a general computer and an ordinary single-pixel detector.
The influences of the speckle patterns, defocusing, and the number of measurements on imaging quality are discussed.
The results show that this computational GI technique has the depth-resolving
imaging capabilities, and it enables image reconstruction with less measurements by adjusting the speckle size.

Furthermore, computation GI with pseudo-randomly patterned illumination from a LCD make to possible to achieve 3D and color
imaging. The presented theoretical framework of structured illumination holds great potential,
such as combining with compressive sensing algorithm, light source coding technology and so on.
Meanwhile, we find that the source of pseudo-randomly distributed patterns in the computer screen is similar to
the true thermal light, which is closer to practical applications, such as single-pixel imaging, light detection and ranging, and fluorescence microscopy.

\begin{center}{\bf ACKNOWLEDGEMENTS}\end{center}

The authors thank the supports from the National Basic Research Program of China under Grant No.2012CB921900, the National Natural Science Foundation
of China (11174084, 10934011).

\end{document}